\pdfoutput=1

\documentclass[12pt]{article}
\usepackage{epsfig}

\def\beq{\begin{equation}}
\def\eeq#1{\label{#1}\end{equation}}
\def\eeqn{\end{equation}}

\def\beqa{\begin{eqnarray}}
\def\eeqa#1{\label{#1}\end{eqnarray}}
\def\eeqan{\end{eqnarray}}

\let\bar=\overbar

\def\etal{{\it et al.}}
\def\ie{{\it i.e.}}

\def\D{{\cal D}}

\def\Dslash{\not{\hbox{\kern-4pt $D$}}}
\def\dslash{\not{\hbox{\kern-2pt $\del$}}}

\def\msb{{\bar{\ssstyle M \kern -1pt S}}}

\def\eps{\epsilon}

\def\Title#1{\begin{center} {\Large {\bf #1} } \end{center}}

\usepackage{ifthen} 
\newboolean{pdflatex}
\setboolean{pdflatex}{true} 

\newboolean{articletitles}
\setboolean{articletitles}{true} 

\newboolean{uprightparticles}
\setboolean{uprightparticles}{false} 

\usepackage{lineno}  
\usepackage{xspace} 

\usepackage{graphicx}  
\usepackage{color}
\usepackage{colortbl}
\graphicspath{{./figs/}} 

\usepackage{amsmath} 
\usepackage{amssymb}
\usepackage{amsfonts}
\usepackage{upgreek} 

\newcommand*\patchAmsMathEnvironmentForLineno[1]{%
\expandafter\let\csname old#1\expandafter\endcsname\csname #1\endcsname
\expandafter\let\csname oldend#1\expandafter\endcsname\csname
end#1\endcsname
 \renewenvironment{#1}%
   {\linenomath\csname old#1\endcsname}%
   {\csname oldend#1\endcsname\endlinenomath}%
}
\newcommand*\patchBothAmsMathEnvironmentsForLineno[1]{%
  \patchAmsMathEnvironmentForLineno{#1}%
  \patchAmsMathEnvironmentForLineno{#1*}%
}
\AtBeginDocument{%
\patchBothAmsMathEnvironmentsForLineno{equation}%
\patchBothAmsMathEnvironmentsForLineno{align}%
\patchBothAmsMathEnvironmentsForLineno{flalign}%
\patchBothAmsMathEnvironmentsForLineno{alignat}%
\patchBothAmsMathEnvironmentsForLineno{gather}%
\patchBothAmsMathEnvironmentsForLineno{multline}%
}

\usepackage{hyperref}    
\usepackage[all]{hypcap} 

\def\lhcb {\mbox{LHCb}\xspace}
\def\ux85 {\mbox{UX85}\xspace}

\ifthenelse{\boolean{uprightparticles}}%
{

 \def\Pmu         {\ensuremath{\upmu}\xspace}

 \def\Ppi         {\ensuremath{\uppi}\xspace}

 \def\Pphi        {\ensuremath{\upphi (1020)}\xspace}

 \def\Ppsi        {\ensuremath{\uppsi}\xspace}

 \def\PDelta      {\ensuremath{\Delta}\xspace}                 
 \def\PXi      {\ensuremath{\Xi}\xspace}                 
 \def\PLambda      {\ensuremath{\Lambda}\xspace}                 
 \def\PSigma      {\ensuremath{\Sigma}\xspace}                 
 \def\POmega      {\ensuremath{\Omega}\xspace}                 
 \def\PUpsilon      {\ensuremath{\Upsilon}\xspace}                 
 
 \def\PB      {\ensuremath{\mathrm{B}}\xspace}                 
                  
 \def\PD      {\ensuremath{\mathrm{D}}\xspace}

 \def\PJ      {\ensuremath{\mathrm{J}}\xspace}                 
 \def\PK      {\ensuremath{\mathrm{K}}\xspace}

 \def\Pb      {\ensuremath{\mathrm{b}}\xspace}                 
 \def\Pc      {\ensuremath{\mathrm{c}}\xspace}

 \def\Pi      {\ensuremath{\mathrm{i}}\xspace}

 \def\Ps      {\ensuremath{\mathrm{s}}\xspace}                 
 \def\Pt      {\ensuremath{\mathrm{t}}\xspace}

}
{

 \def\Pmu         {\ensuremath{\mu}\xspace}

 \def\Ppi         {\ensuremath{\pi}\xspace}

 \def\Pphi        {\ensuremath{\phi}\xspace}

 \def\Ppsi        {\ensuremath{\psi}\xspace}                 
                  
 \mathchardef\PDelta="7101
 \mathchardef\PXi="7104
 \mathchardef\PLambda="7103
 \mathchardef\PSigma="7106
 \mathchardef\POmega="710A
 \mathchardef\PUpsilon="7107
                  
 \def\PB      {\ensuremath{B}\xspace}                 
                  
 \def\PD      {\ensuremath{D}\xspace}

 \def\PJ      {\ensuremath{J}\xspace}                 
 \def\PK      {\ensuremath{K}\xspace}

 \def\Pb      {\ensuremath{b}\xspace}                 
 \def\Pc      {\ensuremath{c}\xspace}

 \def\Pi      {\ensuremath{i}\xspace}

 \def\Ps      {\ensuremath{s}\xspace}                 
 \def\Pt      {\ensuremath{t}\xspace}

}

\def\mup        {\ensuremath{\Pmu^+}\xspace}
\def\mun        {\ensuremath{\Pmu^-}\xspace} 

\def\squark    {\ensuremath{\Ps}\xspace}

\def\cquark    {\ensuremath{\Pc}\xspace}
\def\cquarkbar {\ensuremath{\overline \cquark}\xspace}

\def\bquark    {\ensuremath{\Pb}\xspace}

\def\tquark    {\ensuremath{\Pt}\xspace}

\def\pion  {\ensuremath{\Ppi}\xspace}

\def\pip   {\ensuremath{\pion^+}\xspace}

\def\pipi  {\ensuremath{\pion^+\pion^-}\xspace}

\def\kaon  {\ensuremath{\PK}\xspace}
  \def\Kbar  {\kern 0.2em\overline{\kern -0.2em \PK}{}\xspace}

\def\Kz    {\ensuremath{\kaon^0}\xspace}
\def\Kzb   {\ensuremath{\Kbar^0}\xspace}
\def\KzKzb {\ensuremath{\Kz \kern -0.16em \Kzb}\xspace}
\def\Kp    {\ensuremath{\kaon^+}\xspace}
\def\Km    {\ensuremath{\kaon^-}\xspace}
\def\Kmpip {\ensuremath{\kaon^-\pion^+}\xspace}

\def\KpKm  {\ensuremath{\Kp \kern -0.16em \Km}\xspace}

\def\Kstarzb {\ensuremath{\Kbar^{*0}}\xspace}

\def\Kstarzbm {\ensuremath{\Kbar^{*0}(892)}\xspace}

\def\KK      {\ensuremath{\kaon^{+}\kaon^{-}}\xspace}

  \def\Dbar    {\kern 0.2em\overline{\kern -0.2em \PD}{}\xspace}
\def\D       {\ensuremath{\PD}\xspace}

\def\Dz      {\ensuremath{\D^0}\xspace}
\def\Dzb     {\ensuremath{\Dbar^0}\xspace}
\def\DzDzb   {\ensuremath{\Dz {\kern -0.16em \Dzb}}\xspace}
\def\Dp      {\ensuremath{\D^+}\xspace}
\def\Dm      {\ensuremath{\D^-}\xspace}

\def\DpDm    {\ensuremath{\Dp {\kern -0.16em \Dm}}\xspace}

\def\B       {\ensuremath{\PB}\xspace}
  \def\Bbar    {\kern 0.18em\overline{\kern -0.18em \PB}{}\xspace}

\def\Bu      {\ensuremath{\B^+}\xspace}

\def\Bs      {\ensuremath{\B^0_\squark}\xspace}
\def\Bsb     {\ensuremath{\Bbar^0_\squark}\xspace}

\def\jpsi     {\ensuremath{{\PJ\mskip -3mu/\mskip -2mu\Ppsi\mskip 2mu}}\xspace}

  \def\Y#1S{\ensuremath{\PUpsilon{(#1S)}}\xspace}

\def\Lbar {\ensuremath{\kern 0.1em\overline{\kern -0.1em\PLambda}}\xspace}

\newcommand{\decay}[2]{\ensuremath{#1\!\to #2}\xspace}         

\def\to                 {\ensuremath{\rightarrow}\xspace}

\def\eps   {\ensuremath{\varepsilon}\xspace}

\def\CP                {\ensuremath{C\!P}\xspace}

\def\Vcs  {\ensuremath{V_{\cquark\squark}}\xspace}
\def\Vts  {\ensuremath{V_{\tquark\squark}}\xspace}

\def\Vcb  {\ensuremath{V_{\cquark\bquark}}\xspace}
\def\Vtb  {\ensuremath{V_{\tquark\bquark}}\xspace}

\newcommand{\dms}{\ensuremath{\Delta m_{\squark}}\xspace}

\newcommand{\DG}{\ensuremath{\Delta\Gamma}\xspace}
\newcommand{\DGs}{\ensuremath{\Delta\Gamma_{\squark}}\xspace}

\newcommand{\Gs}{\ensuremath{\Gamma_{\squark}}\xspace}

\newcommand{\phis}{\ensuremath{\phi_{\squark}}\xspace}

\def\BsToJPsiPhi  {\decay{\Bs}{\jpsi\phi}}
\def\BsToJPsiKK   {\decay{\Bs}{\jpsi \Kp\Km}}
\def\BsToJPsipipi {\decay{\Bs}{\jpsi \pipi}}

\def\BsToJPsiKst  {\decay{\Bs}{\jpsi\Kstarzb}}
\def\BsToJPsiKpi  {\decay{\Bs}{\jpsi\Kmpip}}
\def\BsToJPsiKstm  {\decay{\Bs}{\jpsi\Kstarzbm}}
\def\BuToJPsiK  {\decay{\Bu}{\jpsi\Kp}}

\def\Psimumu      {\decay{\jpsi}{\mup\mun}}

\def\AT#1     {\ensuremath{A_{\mathrm{T}}^{#1}}\xspace}           

\def\C#1      {\ensuremath{\mathcal{C}_{#1}}\xspace}                       
\def\Cp#1     {\ensuremath{\mathcal{C}_{#1}^{'}}\xspace}                    
\def\Ceff#1   {\ensuremath{\mathcal{C}_{#1}^{\mathrm{(eff)}}}\xspace}        
\def\Cpeff#1  {\ensuremath{\mathcal{C}_{#1}^{'\mathrm{(eff)}}}\xspace}       
\def\Ope#1    {\ensuremath{\mathcal{O}_{#1}}\xspace}                       
\def\Opep#1   {\ensuremath{\mathcal{O}_{#1}^{'}}\xspace}                    

\newcommand{\tev}{\ensuremath{\mathrm{\,Te\kern -0.1em V}}\xspace}
\newcommand{\gev}{\ensuremath{\mathrm{\,Ge\kern -0.1em V}}\xspace}
\newcommand{\mev}{\ensuremath{\mathrm{\,Me\kern -0.1em V}}\xspace}
\newcommand{\kev}{\ensuremath{\mathrm{\,ke\kern -0.1em V}}\xspace}
\newcommand{\ev}{\ensuremath{\mathrm{\,e\kern -0.1em V}}\xspace}
\newcommand{\gevc}{\ensuremath{{\mathrm{\,Ge\kern -0.1em V\!/}c}}\xspace}
\newcommand{\mevc}{\ensuremath{{\mathrm{\,Me\kern -0.1em V\!/}c}}\xspace}
\newcommand{\gevcc}{\ensuremath{{\mathrm{\,Ge\kern -0.1em V\!/}c^2}}\xspace}
\newcommand{\gevgevcccc}{\ensuremath{{\mathrm{\,Ge\kern -0.1em V^2\!/}c^4}}\xspace}
\newcommand{\mevcc}{\ensuremath{{\mathrm{\,Me\kern -0.1em V\!/}c^2}}\xspace}

\def\invfb   {\ensuremath{\mbox{\,fb}^{-1}}\xspace}

\def\ps   {\ensuremath{{\rm \,ps}}\xspace}
\def\fs   {\ensuremath{\rm \,fs}\xspace}

\def\invps{\ensuremath{{\rm \,ps^{-1}}}\xspace}

\def\gsim{{~\raise.15em\hbox{$>$}\kern-.85em
          \lower.35em\hbox{$\sim$}~}\xspace}
\def\lsim{{~\raise.15em\hbox{$<$}\kern-.85em
          \lower.35em\hbox{$\sim$}~}\xspace}

\def\rad{\ensuremath{\rm \,rad}\xspace}

\def\tell1  {TELL1\xspace}
\def\ukl1   {UKL1\xspace}

\usepackage{cite} 
\usepackage{mciteplus}

\begin{document}

\Title{Measurement of the \CP violation phase $\phis$ in the $B_s$ system at \lhcb}

\bigskip\bigskip

\begin{raggedright}  

{\it Gerhard Raven\index{Raven, G.},\\
on behalf of the LHCb Collaboration\\
VU University Amsterdam \& Nikhef \\
Science Park 105 \\
1098 XG Amsterdam, THE NETHERLANDS}\\
\bigskip
Proceedings of the CKM 2012, the 7$^\mathrm{th}$ International Workshop on the CKM Unitarity Triangle, University of Cincinnati, USA, 28 September - 2 October 2012
\bigskip\bigskip
\end{raggedright}

\section{Introduction}

The interference between \Bs\ decay amplitudes to \CP eigenstates $\jpsi X$ directly or including $\Bs-\Bsb$ oscillation
gives rise to a measurable \CP violating phase \phis.
In the Standard Model (SM), these decays are dominated by the tree level $b\to c (\overline{c}s)$ transition, and hence
$\phis$ is predicted to be $\phi_s^{SM}\approx - 2\beta_s$, where
$\beta_s = \arg\left(-\Vts\Vtb^*/\Vcs\Vcb^*\right)$\cite{SMPHENO}. The indirect determination, via global
fits to experimental data, predicts $2\beta_s = 0.036 \pm 0.002 \rad$\cite{SMFIT}. However, non-SM 
constributions to  \Bs--\Bsb\ mixing may alter this prediction\cite{NP}.

Here, a preliminary measurement of \phis with \BsToJPsiKK decays\cite{LHCb-CONF-2012-002}, and the measurement using 
\BsToJPsipipi decays\cite{arXiv:1204.5675}, both obtained from a sample of  1\invfb of $pp$ collisions, collected by the \lhcb experiment 
at a centre-of-mass energy $\sqrt{s}=7\tev$ during 2011, are presented.
In addition, the determination of the branching fraction and polarization fractions of \BsToJPsiKst\cite{arXiv:1208.0738},
which can be used to constrain the contribution of sub-leading penguin contributions to the decay \BsToJPsiPhi, is shown.

\section{\BsToJPsiKK}

The phenomenological aspects of \BsToJPsiKK  decays, where the \KK pair arises through the \Pphi\ resonance, are 
described in many articles, e.g.\ in Refs~\cite{PHENOII}.
The effects induced by the sub-leading penguins contribution are discussed, for example, in~\cite{PenguinTheory}.

As the   \BsToJPsiPhi{}  decay proceeds via two intermediate spin-1 particles (\ie\ with the \KK\  pair
in a P-wave), the final state is a superposition of \CP-even and \CP-odd states depending upon the relative orbital angular momentum
between the \jpsi and the $\phi$.  In addition, the same final state can be produced with \KK\ pairs with zero relative orbital angular
momentum (S-wave)~\cite{Stone:2008ak}, which produces a \CP-odd contribution.  In order to measure $\phis$ it is necessary to disentangle
these \CP-even and \CP-odd components. This is achieved by analysing the distribution of the reconstructed decay angles.

\begin{figure}[htb]
\begin{center}
\epsfig{file=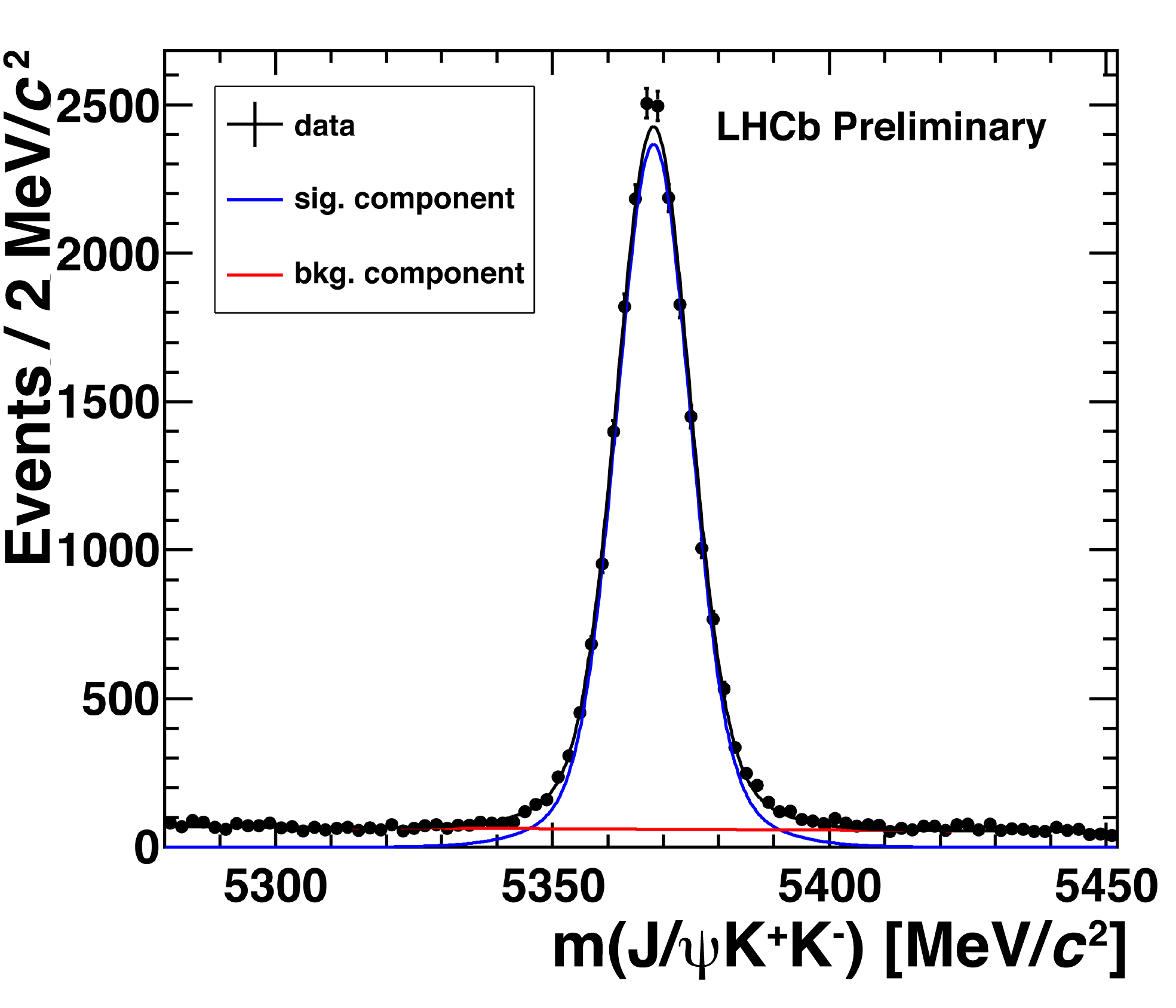,height=2.3in}
\epsfig{file=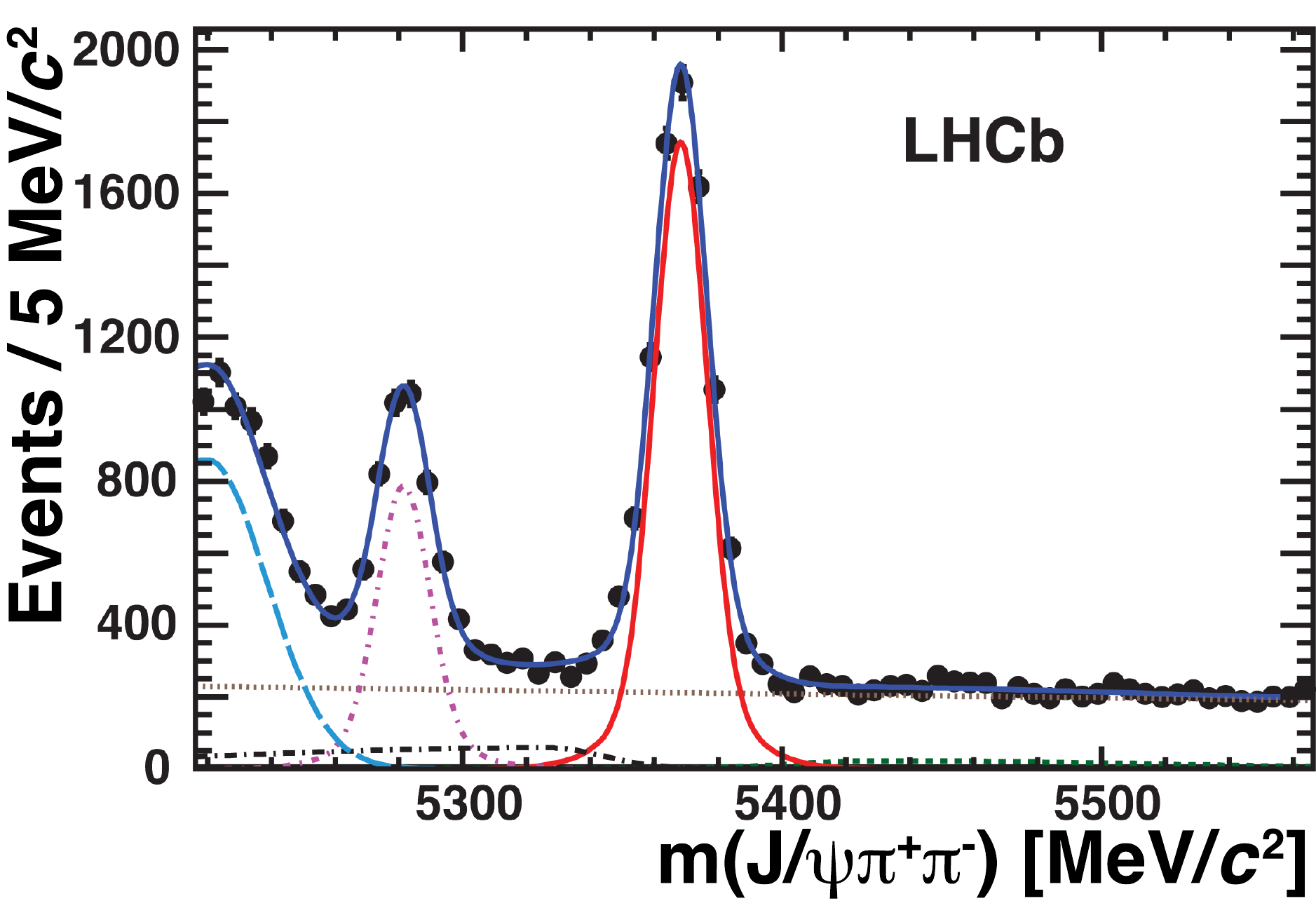,height=2.2in}
\caption{ Left: Invariant mass distribution of selected \BsToJPsiKK\ candidates. 
The background is shown as the horizontal (red) line. Right: Invariant mass distribution of selected 
\BsToJPsipipi candidates. The signal is shown as the red solid line. 
Backgrounds are combintorial (brown dotted), \BsToJPsiPhi (black long-dot) and \BuToJPsiK,\jpsi\pip (dashed green).
Additional backgrounds are shown, but are irrelevant as the analysis only uses the data 
above a mass of 5346 \mevcc.
}
\label{fig:invariantmasses}
\end{center}
\end{figure}
The event selection is the same as the one  described in \cite{arXiv:1112.3183}. 
However, due to the increased instantaneous luminosity, the trigger conditions were changed in the second half of the 2011 data 
taking period, and decay time biasing cuts were added. A non-parametric description of this acceptance is determined using events triggered by 
a prescaled trigger line without these extra cuts, and accounted for in the fit. An additional correction, based on simulation, is 
made to correct for the small reduction of tracking efficiency for tracks originating from vertices far from the beamline.
The decay angle acceptance is obtained from simulation, and the difference between the observed and simulated kaon momentum spectra
is used to derive the corresponding systematic uncertainty.

To account for the decay time resolution of the detector, the Probability Density Functions (PDFs) used in the fit are 
convolved with a Gaussian whose width is proportional to the per-event computed decay time resolution. 
The scale factor is determined from the \Psimumu\ component of the prompt background 
and is found to be $1.45 \pm 0.06$, where the error includes both statistical and systematic uncertainties, the latter 
derived from simulation. The scale factor is allowed to vary within its uncertainty in the fit. The effective average decay 
time resolution is $45 \pm 2 \fs$. 

The initial flavour of the signal decay is inferred from the other $b$-hadron in the event by the opposite-side flavour tagger, described in \cite{arXiv:1202.4979}.
This algorithm combines  muons, electrons and kaons with large transverse momentum, and the charge of inclusively 
reconstructed secondary vertices and provides an estimated per-event mistag probability, which is calibrated with \BuToJPsiK\  decays. 
The effective average mistag probability $w = (36.8 \pm  0.2 \pm 0.7)\%$ yields, when combined with the efficiency $\eps_{\mathrm{tag}} = ( 33.0 \pm 0.3 )\%$, 
an effective tagging power of $ Q = \eps_{\mathrm{tag}} (1-2w)^2 = (2.29 \pm 0.07 \pm 0.26 ) \%$.
The effects of a possible difference in mistag probability between \Bs\ and \Bsb, and of a potential 
tagging efficiency asymmetry were estimated to be negligible. The uncertainties from the flavour tag calibration 
are included by allowing the calibration parameters to vary in the likelihood fit within their uncertainties.

\section{\BsToJPsipipi}

The analysis of the \BsToJPsipipi\ channel is published as \cite{arXiv:1204.5675}. It is also based upon the 
1\invfb of data taken during 2011, and utilizes a \pipi invariant mass range of
 $[775-1550]$~\mevcc. Although this range is dominated by $f_0(980)$, it also encompasses $f_2(1270)$, $f_0(1370)$ and
a non-resonant component. It has has been shown \cite{arXiv:1204.5643} that this range is almost entirely \CP-odd, with
a \CP-odd fraction $> 97.7\%$ at 95\% C.L.  As a result, no angular analysis is required.
The analysis relies upon the same tagging information as the \BsToJPsiKK\ analysis. 
The invariant mass distribution of the selected events is shown on the right of Fig. \ref{fig:invariantmasses}.

\section{Results}

The \CP violating phase \phis is extracted from the \BsToJPsiKK\ sample with an unbinned maximum likelihood fit 
to the candidate invariant mass, decay time, initial \Bs flavour and the decay angles.
The signal and background PDFs of the likelihood are given in \cite{arXiv:1112.3183}. We determine several physics parameters,
namely the decay width, \Gs, the decay width difference between the light and heavy \Bs mass eigenstate 
\DG, and the polarization amplitudes of the \KK\ system. In the fit we parameterise the P-wave transversity amplitudes, 
$A_i$, by their absolute value at production time, $|A_i(0)|$, and their phases $\delta_i$ 
and adopt the convention $\delta_0=0$. We utilize the following normalization: $|A_0(0)|^2 + |A_\parallel(0)|^2 + |A_\perp(0)|^2 = 1$,
and define the fraction of S-wave contribution $F_S = |A_S(0)|^2/(1 + |A_S(0)|^2)$.
This choice of the normalization insures that the P-wave amplitudes have the same value independently of the range of the \KK\ invariant mass chosen. 
We use the measurement of the \Bs\ oscillation frequency $\dms = 17.63 \pm± 0.11\ps^{-1}$ \cite{arXiv:1112.4311}
and allow it to vary within its uncertainty.  The values obtained, as well as their statistical and systematic uncertainties, 
are given in Table 1.  All parameters, except $\delta_\parallel$, have a well behaved parabolic profile likelihood.
The exception for the $\delta_\parallel$ is caused by the fact that its central value lies just above $\pi$ which implies 
that it is almost degenerate with the ambiguous solution at $\delta_\parallel\to2\pi-\delta_\parallel$ which lies symmetrically just below 
$\pi$.  The quoted 68\% C.L. encompasses both minima.

\begin{table}[htb]
\begin{center}
\begin{tabular}{|l|c|c|c|}  
\hline
Parameter & Value & Stat. & Syst. \\
\hline
\hline
\Gs              &    0.6580 & 0.0054 & 0.0066 \\
\DGs             &    0.116  & 0.018  & 0.006 \\
$|A_\perp(0)|^2$ &    0.246  & 0.010  & 0.013 \\
$|A_0(0)|^2$     &    0.523  & 0.007  & 0.024 \\
$F_S$            &    0.022  & 0.012  & 0.007 \\
$\delta_\perp$   &    2.90   & 0.36   & 0.07  \\
$\delta_\parallel$ &  \multicolumn{2}{|c|}{[2.81,3.47]}  & 0.13    \\
$\delta_S$         &  2.90   & 0.36   & 0.08  \\
\phis              &  -0.001 & 0.101  & 0.027 \\
\hline
\end{tabular}
\caption{Results for the physics parameters and their statistical and systematic uncertainties. 
We quote a 68\% C.L. interval for $\delta_\parallel$ as described in the text.}
\label{tab:result}
\end{center}
\end{table}
\begin{figure}[htb]
\begin{center}
\epsfig{file=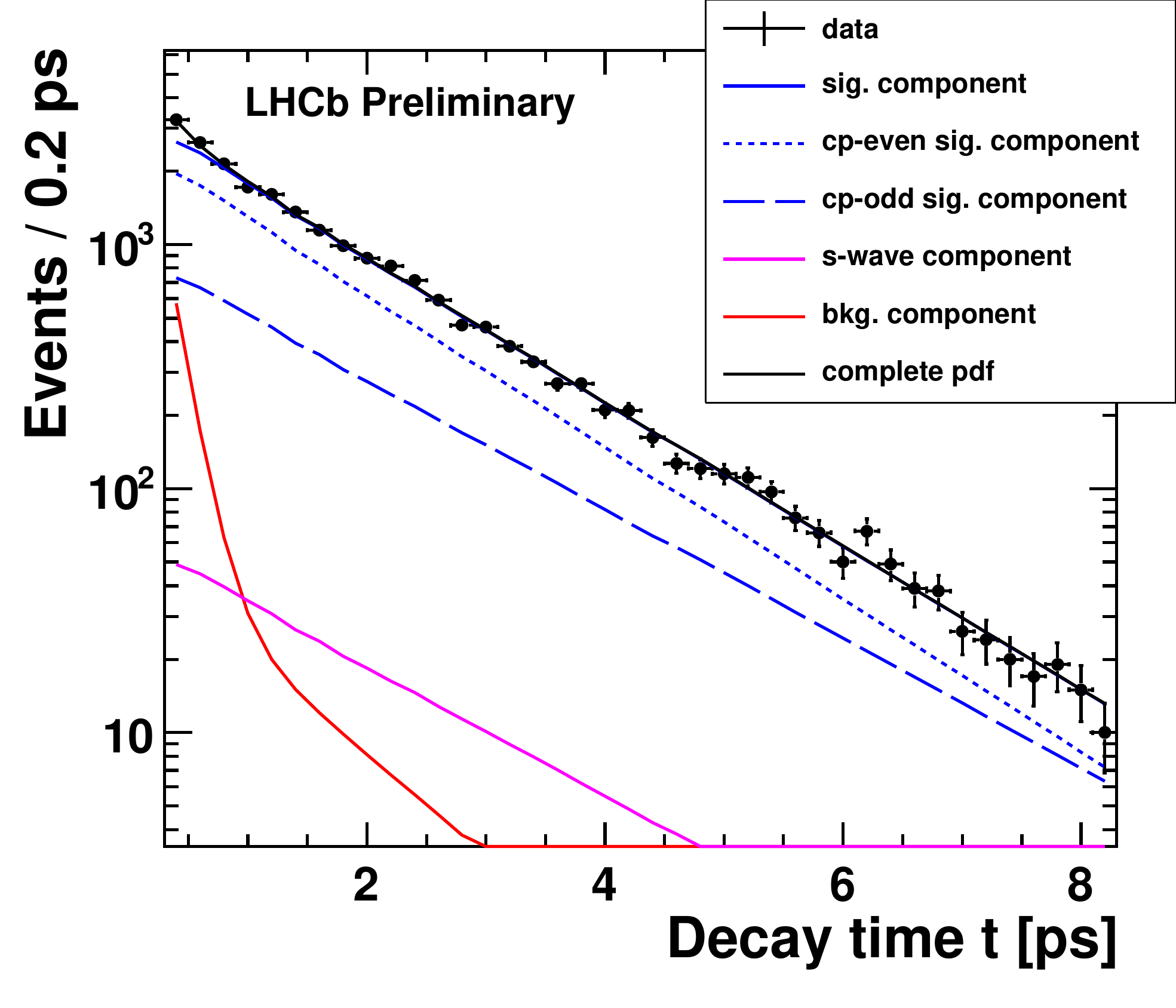,height=2.6in}
\epsfig{file=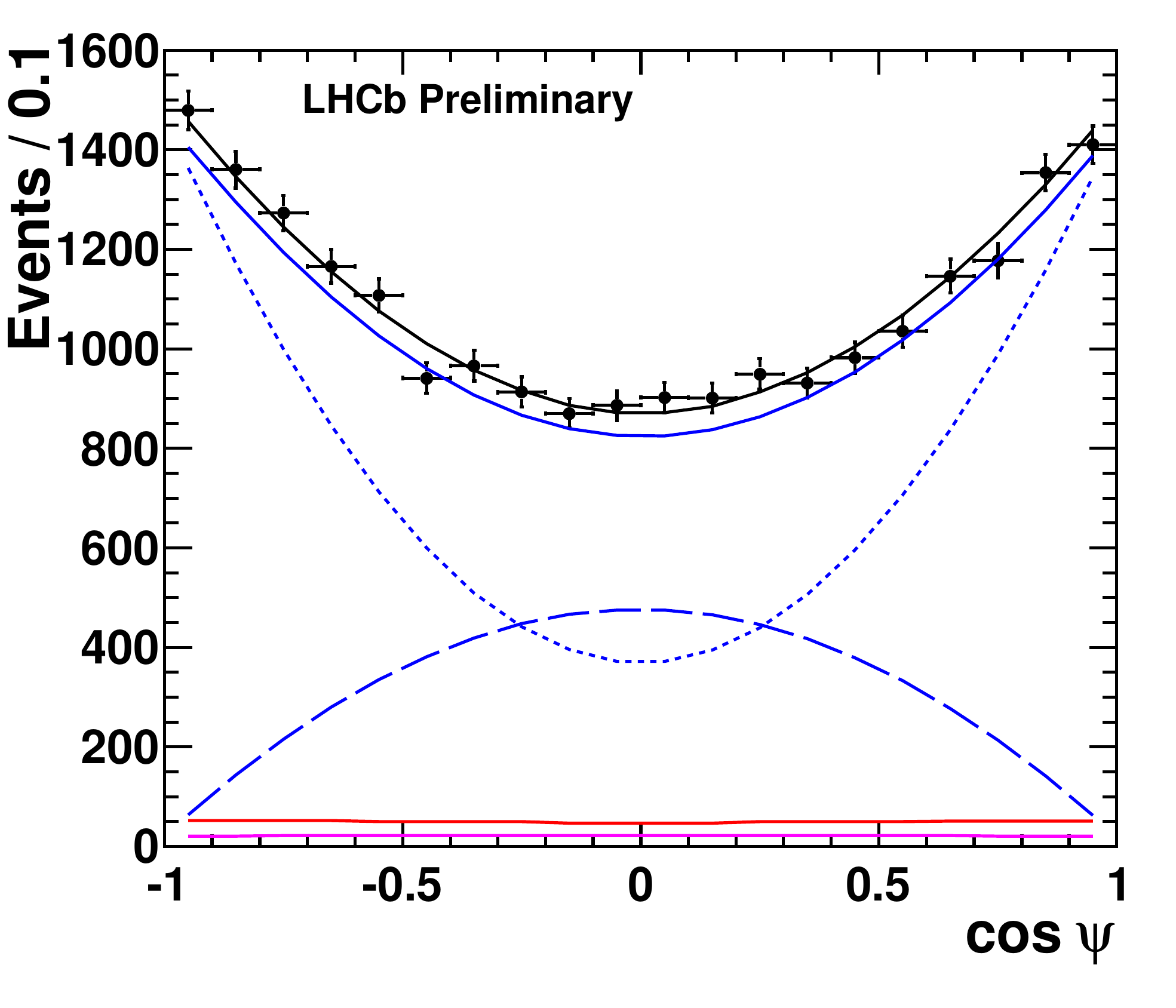,height=2.6in}
\epsfig{file=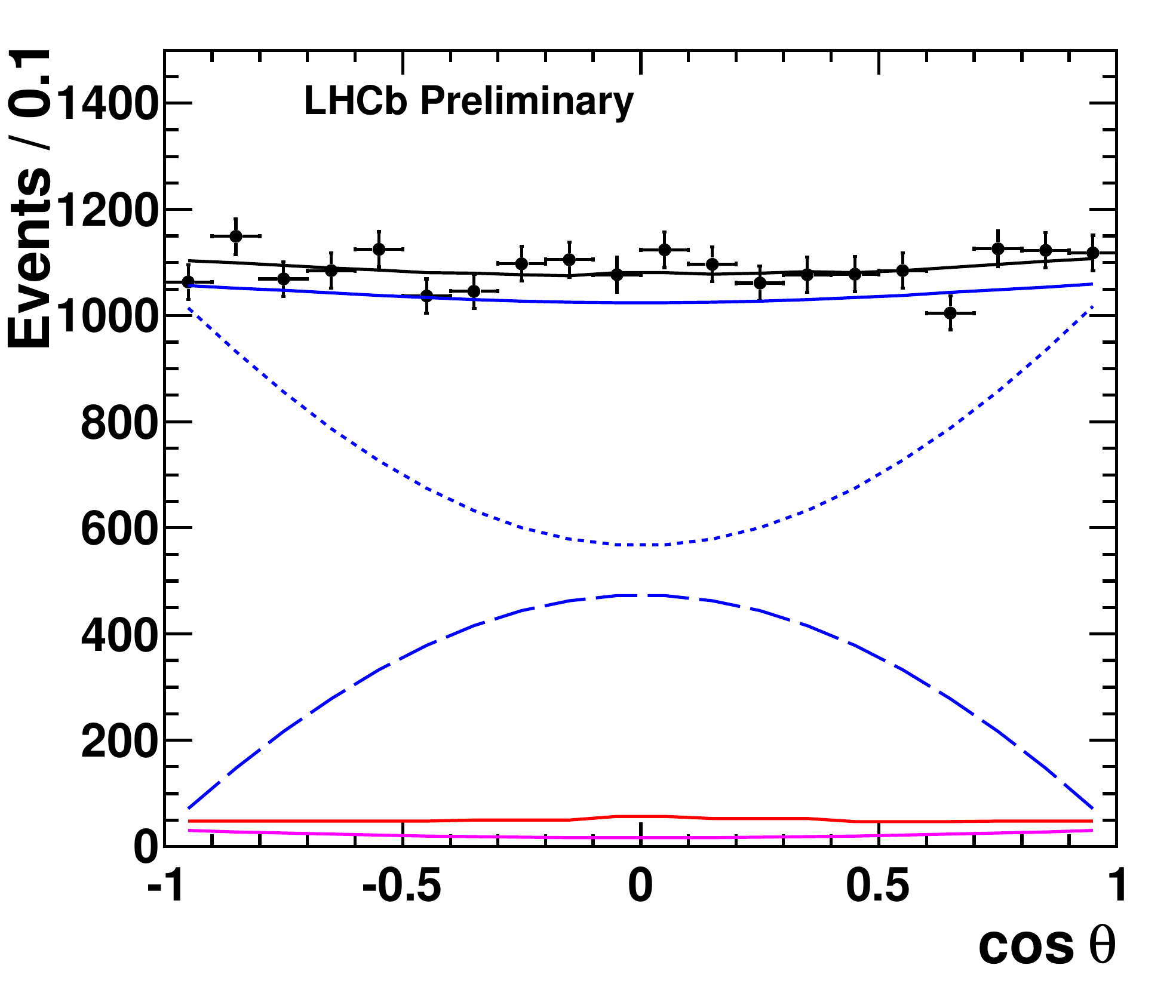,height=2.6in}
\epsfig{file=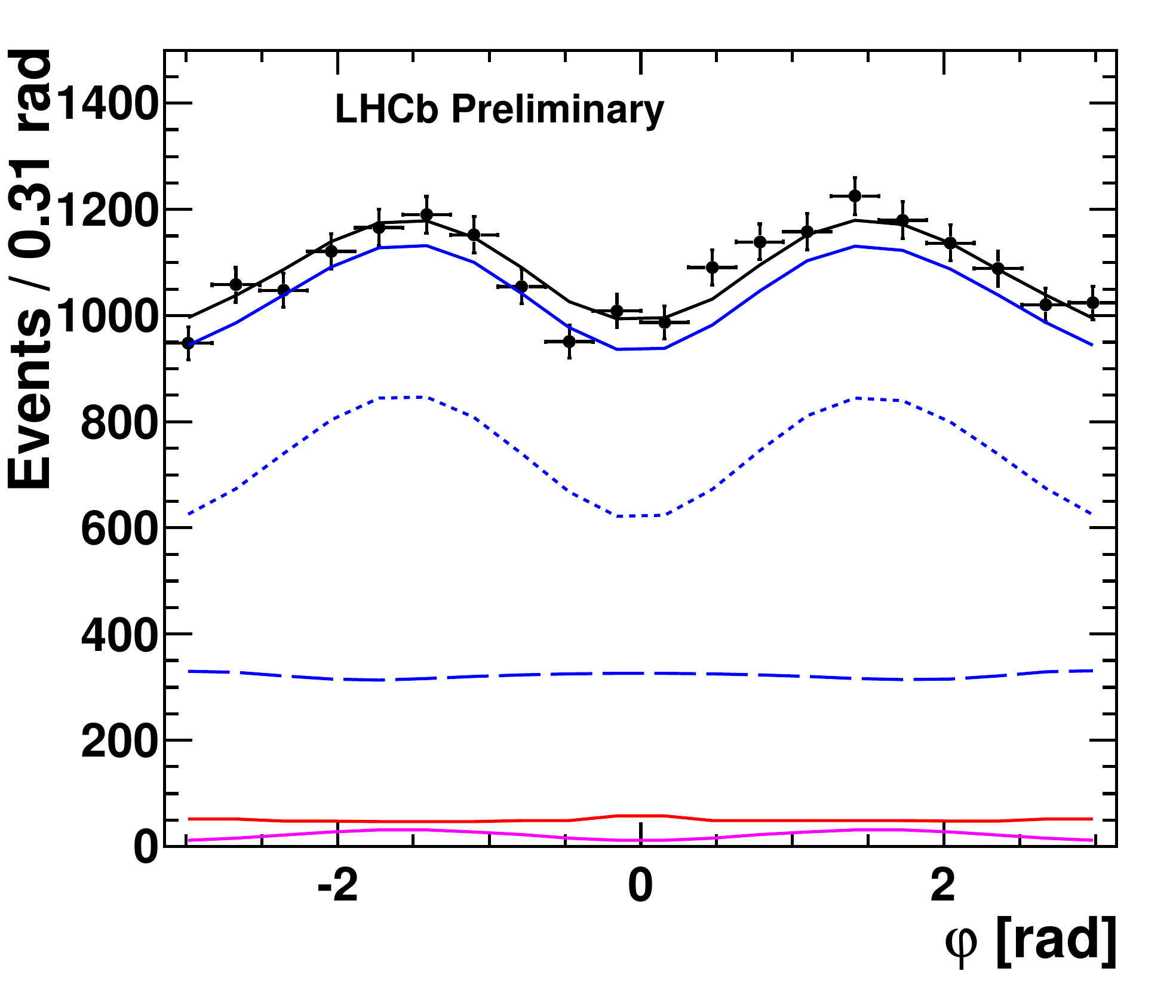,height=2.6in}
\caption{ Data points overlaid with fit projections for the decay time and transversity angle distributions 
in a mass range of $\pm 20 \mevcc$ around the reconstructed \Bs\ mass. The total fit result is represented 
by the black line. The signal component is represented by the solid blue line; the dashed and dotted blue lines 
show the \CP-odd and \CP-even signal components respectively. The S-wave component is represented by the solid 
pink line. The background component is given by the red line.
}
\label{fig:jpsiphiprojections}
\end{center}
\end{figure}

The results for \phis\ and \DGs\ are in good agreement with the Standard Model predictions\cite{arXiv:1106.4041}. 
Figure \ref{fig:jpsiphiprojections}  shows the projection of the fitted PDF on the decay time and 
the transversity angle distributions for candidates with an invariant mass within $20 \mevcc$ around the nominal \Bs\ mass. 
The systematic uncertainties quoted in Table 1 are those which are not directly treated in the maximum likelihood fit. 
The systematic uncertainty on \phis\ is dominated by imperfect knowledge of the angular acceptances 
and neglecting potential contributions of direct \CP-violation. 

The \CP violating phase \phis\ is extracted from the \BsToJPsipipi sample  with an unbinned maximum likelihood fit 
to mass, decay time and flavour tag. The result is $\phis = -0.02 \pm 0.17 \pm 0.004$ rad.
The systematic uncertainties from tagging and resolution are included in a similar way as for the \BsToJPsiKK\ analysis. 
Full details can be found in Ref. \cite{arXiv:1204.5675}.
The results from the \BsToJPsiKK\ and \BsToJPsipipi samples are compatible, and, when combined in a simultaneous 
fit, give $\phis = -0.002 \pm 0.083 \pm 0.027$ rad.

This analysis results in a twofold ambiguity $(\phis,\DGs)\leftrightarrow(\pi-\phis,-\DGs)$.
The ambiguity has been resolved \cite{arXiv:1202.4717} using the measurement of the evolution of the relative phase between the \KK P-wave 
and S-wave amplitudes as a function of the \KK mass. The solution with $\DGs>0$ is favoured and only this solution is quoted here.

\section{ \BsToJPsiKpi }

Although the decay \BsToJPsiKK is dominated by the tree level $\bquark\rightarrow \cquark(\cquarkbar \squark)$ 
transition, there are contributions from higher order (penguin) $\bquark\rightarrow \squark (\cquark\cquarkbar)$ processes.
These cannot be calculated reliably and could 
affect the measured asymmetries. It has been suggested that these effects can be controlled
by means of an analysis of the decay $\BsToJPsiKst$, where the penguin diagrams are
not suppressed relative to the tree level, and $SU(3)$ flavour symmetry can be used to
determine the relevant hadronic parameters~\cite{PenguinTheory}.

LHCb has searched\cite{arXiv:1208.0738} for the decay \BsToJPsiKst with $\Kstarzb \to \Kmpip$ with
\mbox{$0.37$ fb$^{-1}$} of $pp$ collisions at $\sqrt{s}$ = 7~\tev, and observes $114 \pm 11$ \BsToJPsiKpi\ signal candidates. 
The \Kmpip mass spectrum of the candidates in the \Bs peak is dominated by the \Kstarzbm contribution. 
Subtracting the non-resonant \Kmpip component, the branching fraction of \BsToJPsiKstm is $\left(
4.4_{-0.4}^{+0.5} \pm 0.8 \right) \times 10^{-5}$. 
A fit to the angular distribution yields the 
polarization fractions $f_L = 0.50 \pm 0.08 \pm 0.02$ and $f_{\parallel} = 0.19^{+0.10}_{-0.08} \pm
0.02$.


\section{Conclusion}

We have performed a time-dependent angular analysis of approximately 21,200 
\BsToJPsiKK decays obtained from 1\invfb of $pp$ collisions at $\sqrt{s}=7\tev$
collected during 2011. From these events we extract:
\begin{eqnarray*}
  \phis &=& -0.001 \pm 0.101 (\mathrm{stat}) \pm 0.027 (\mathrm{syst}) \rad, \\
  \Gs &=& 0.658  \pm 0.005 (\mathrm{stat}) \pm 0.007 (\mathrm{syst}) \invps, \\
  \DGs &=& 0.116  \pm 0.018 (\mathrm{stat}) \pm 0.006 (\mathrm{syst}) \invps. \\
\end{eqnarray*}
When combined with the result from an independent analysis of approximately 7421 \BsToJPsipipi\ decays,
we find
\begin{equation}
  \phis = -0.002 \pm 0.083 (\mathrm{stat}) \pm 0.027 (\mathrm{syst}) \rad.
\end{equation}
This is the world's most precise measurement of \phis and the first direct observation for a non-zero
value for \DGs. These results are in good agreement with Standard Model predictions \cite{arXiv:1106.4041}.

\end{document}